\documentstyle[11pt,newpasp,twoside,epsf]{article}
\def\edcomment#1{\iffalse\marginpar{\raggedright\sl#1\/}\else\relax\fi}
\reversemarginpar

\begin{document}
\title{Braking and bouncing}

\author{U. Kolb} 

\affil{Department of Physics and Astronomy, The Open University,
\\Walton Hall, Milton Keynes MK7 6AA, UK}

\begin{abstract}
I discuss the case for an efficient orbital braking mechanism in CVs,
and the effect of nova outbursts on the observed range of mass 
transfer rates. I review the continuing problem that models have in
interpreting the short--period cut--off of the CV period distribution.   
\end{abstract}

\section{Introduction}

The commonly accepted model for the formation of cataclysmic variables 
(CVs) with hydrogen--rich main-sequence or brown dwarf donor stars
involves four main assumptions.
\begin{enumerate}
\item The progenitor system was a wide enough binary to allow the
progenitor star of the white dwarf, an intermediate--mass star, to
evolve to giant dimensions prior to Roche--lobe contact.
\item  The system evolved through a common envelope (CE) phase which
exposes the giant's degenerate core, the future white dwarf, and
tightens the orbit.
\item Orbital angular momentum (AM) losses drive the post-common
envelope system into contact, and drive mass transfer in the
semi--detached system.
\end{enumerate}

There are various clues to the magnitude of the AM loss rate.

The overwhelming majority of CVs below 2 hrs orbital period, $P$, are
dwarf novae, while at longer periods novalikes and dwarf novae are
observed with roughly equal frequency. Dwarf novae are unstable disc
accretors where the mass transfer rate $\dot M$ is below a critical
limit $\dot M_{\rm crit}$ so that the hydrogen--ionisation
instability is not suppressed. 
The relative distribution of novalike CVs and dwarf novae with orbital
period thereferore suggests that the transfer rate $\dot M$ is
significantly smaller than $\dot M_{\rm crit}$ 
at short $P$, while $\dot M$ must be close to $\dot M_{\rm crit}$
at longer $P$. As the critical rate typically scales as
$\dot M_{\rm crit} \propto P^{2}$ (e.g.\ Shafter 1992, and references  
therein), the secular mean transfer rate, and hence the AM loss rate,
must be significantly smaller at short orbital periods ($P \la 2$~hr)
than at longer periods. 

The well known disrupted orbital braking model for the orbital
period gap (Spruit \& Ritter 1983, Rappaport, Verbunt \& Joss 1993) is
an extreme manifestation of this dependence: the AM loss rate drops
essentially discontinuously by more than an order of magnitude,
causing the systems to detach at the upper edge of the gap.
The width of the gap requires that the mass transfer rate at this
point is $\dot M \simeq 1-2 \times 10^{-9}$ M$_{\sun}$ yr$^{-1}$ for
unevolved 
Pop~I donor stars (e.g.\ Stehle, Ritter \& Kolb 1996). This is almost
two orders of magnitude larger than the rate driven by gravitational
radiation (GR). Conversely, the short--period cut--off of the observed
period distribution at 78 min, when interpreted as the minimum period
that coincides with the donor's transition from a main--sequence star
to a brown dwarf (Paczy\'nski 1981), seems to require a braking rate
which is at most a factor of 3 above the value set by GR. 

Constraints on $\dot M$ from estimated absolute magnitudes of CVs
suffer from distance uncertainties, difficulties in determining the
duty cycle of dwarf 
novae, and in mapping visual magnitudes to disc accretion rates. 
The deduced transfer rates are, on average, larger above the gap than
below the gap, and there seems to be a considerable intrinsic spread
of the $\dot M$ values at least for systems above the gap (e.g.\
Figure 9.8 in Warner 1995).   

The spectral type of CV donor stars above the gap is, on average,
later than that of an unevolved main--sequence star in thermal
equilibrium that would fill its Roche lobe at the same orbital period
(e.g.\ Baraffe \& Kolb 2000). In the period range 3--6 hrs this is a
natural consequence if the stars experience mass loss at a rate
consistent with the period gap width. 

Townsley and Bildsten (2002) considered the accretional heating of white
dwarfs. They assert that the observationally deduced white dwarf
temperatures are consistent with  a mean accretion rate of order a few
$\times 10^{-9} M_\odot$~yr$^{-1}$ in CVs above the gap, and $\la
10^{-10} M_\odot$~yr$^{-1}$ below the gap.

The last three clues --- gap width, late spectral types,
accretional heating --- constrain the {\it long--term mean} of the
transfer rate, averaged over typically a thermal time of the donor
star.  

In the following three sections I briefly discuss the status of our
understanding of magnetic braking, the effect of nova outbursts on the
observed mass transfer rates, and the shape of the period distribution
below 2 hrs. A recent review that covers also other applications of
the standard disrupted braking model of CV evolution is given by Kolb
(2001).

\section{Magnetic braking}

It has been suggested that magnetic stellar wind braking, usually
thought to be the dominant AM loss mechanism in CVs, is much
weaker than previously thought (Andronov, 
Pinsonneault \& Sills 2001; Pinsonneault, these proceedings). 
This work represents a long overdue generalisation of the so--called  
Skumanich law for solar--type stars to rapidly rotating stars (in some
cases almost as rapid as CV secondaries) and spectral types other than
G. The AM loss rate $\dot J$ has a Skumanich form $\dot J \propto
\omega^3$ at small values of the rotational angular speed $\omega$,
but appears to saturate for rapid rotators to give $\dot J \propto
\omega$. 

A note of caution applies to this ``new'' braking law. Building on 
the work of Krishnamurti et al.\ (1997), Sills et al.\ (2000), and
others, Andronov et al.\ deduced this AM loss rate from the observed
rotation rates in young open clusters. The procedure involves the
deconvolution  
of various physical mechanisms that all affect the surface
rotation of cluster stars: the pre--main sequence contraction of
stars, the redistribution of angular momentum inside the star,
interaction with protostellar discs, and the rotational braking by
magnetic braking. It is clear, though, that the fastest rotators in
young open clusters do not brake (by far) as fast as the 
original Skumanich law would have suggested. 

It is equally clear that the AM loss rate deduced by Andronov et al.\ 
cannot account for the large accretion rates observed above the
period gap, nor for the degree of thermal disequilibrium the spectral
type evidence seems to suggest. Likewise, the period gap
model would have to be modified because the braking rate is 
continuous over the fully convective boundary. Until 
there is a viable alternative to the current thinking that the gap
arises because systems are detached while they cross the gap region,
and unless all novalike systems represent, in evolutionary terms, 
short--lived states with a high accretion rate, the AM loss rate 
in CVs above 3 hrs orbital period must be much larger than the AM loss
rate Andronov et al.\ derive for single stars. 

Potential mechanisms for additional orbital braking include
circumbinary discs (Spruit \& Taam 2001) and accretion disc winds
(Livio \& Pringle 1994), the latter {\em only} in systems close to
mass transfer instability (King \& Kolb 1995).  
However, such alternative scenarios usually fail to reproduce 
sufficiently sharp period gap boundaries. It will be interesting to
watch the evolution of the observed period distribution; the most
recent list of systems (Downes et al.\ 2001, period as of 1 July 2001)
displays gap boundaries which are less sharply defined than in
earlier samples (e.g., compare Figure~1, left panel, with Fig.~1 in
Kolb 1996).  

\begin{figure}
\plottwo{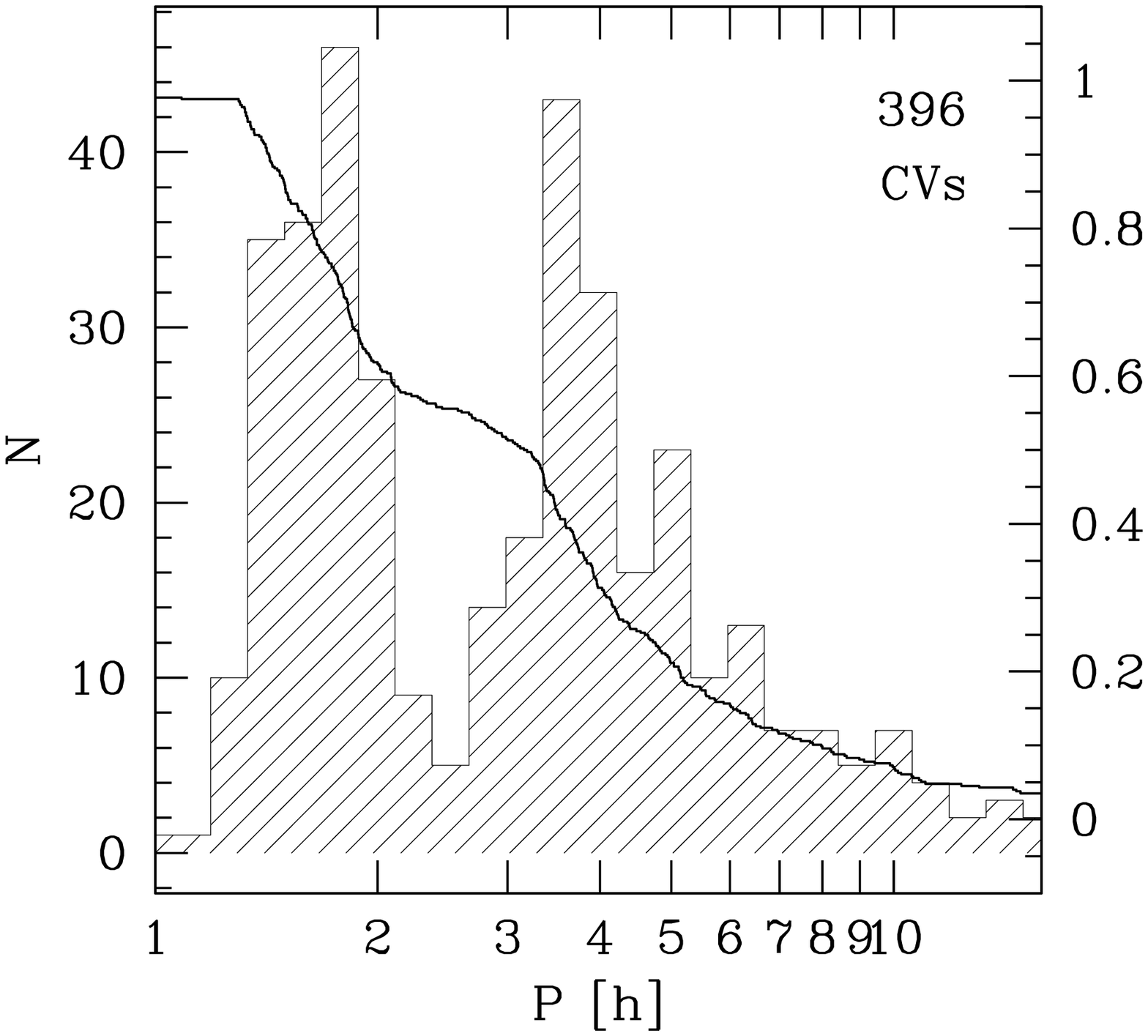}{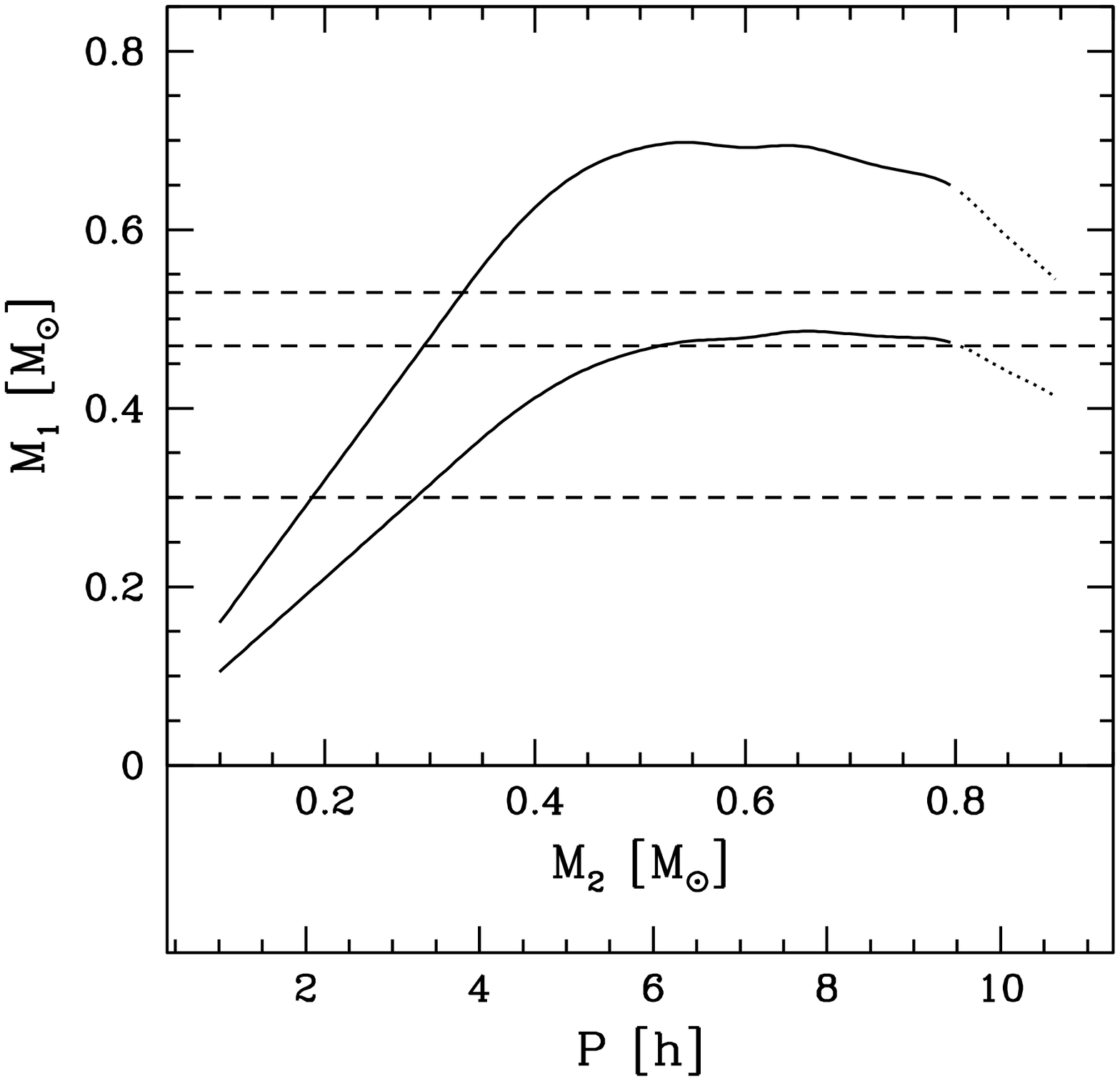}
\caption{{\em Left:} Orbital periods of CVs; data from Downes et al.\
2001, version 1 July 2001. 
{\em Right:} Critical limits in the white dwarf mass ($M_1$) versus
donor mass ($M_2$) plane. Solid: lower limit for conservative mass
transfer (upper curve) and for mass transfer with ballistic ejections
(lower curve). Dashed: lower and upper mass limits for C/O and He
WDs. 
}
\end{figure}

\section{The effect of nova outbursts}

The effect of nova outbursts on the distribution of mass transfer
rates in CVs with similar orbital period (henceforth {\em transfer
rate spectrum} for short) has been revisited by Kolb et al.\
2001. Prior work by Shara et al.\ (1986) pointed out that as a result
of the ejection of a nova shell the transfer rate can drop by one or
more orders of magnitude. They hypothesized that CVs ``hibernate'',
i.e.\ most of the time between nova outbursts they are effectively
detached with no mass transfer, and accrete only in short intervals
that precede a classical nova outburst. 
However, Shara et al.\ (1986) did not consider how the transfer rate
$\dot M$ varies with stellar radius $R$ and Roche lobe radius $R_{\rm
L}$, and how {\em both} of these radii change with time.     

Adopting a simple exponential law,
\begin{displaymath}
\dot M \propto \exp \left( \frac{R-R_{\rm L}}{H}\right),
\end{displaymath}
where $H$ is the photospheric scaleheight (e.g.\ Ritter 1988), and
assuming that the star's thermal relaxation is the same as in the
mean evolution (averaged over individual nova outbursts),  
reveals an area of parameter space where significant widths of
the transfer rate spectra can occur. Generally, this occurs when
the ejected envelope mass $M_{\rm ej}$ is larger than a critical limit,
and the effect is strongest if the ejection is ballistic, i.e.\ if the
affected shell does not gain additional AM while ejected. 
 
Conditions for wide spectra are most favourable if the WD mass ($M_1$)
is small and the system is dynamically unstable against conservative 
mass transfer. For a typical WD mass, $M_1 \simeq 0.6-0.7 M_{\sun}$,
the transfer rate spectrum is wider than 1 dex if  
\begin{equation}
M_{\rm ej} \ga \epsilon \times 3 \times 10^{-4} M_{\sun}.
\end{equation}
The scale factor $\epsilon$, the ratio of photospheric
scaleheight $H$ to stellar radius $R$ in units of $10^{-4}$, is about 
unity for CV donor stars. 
Nova ejecta masses of this order have been suggested both by
theoretical model calculations (e.g.\ Prialnik \& Kovetz 1995) and by 
observational estimates (e.g.\ Starrfield 1999).

Systems that are dynamically unstable in the absence of nova
outbursts do not develop the instability as the transfer rate is kept
in check through nova outbursts.
The phase space of systems that are dynamically unstable in the
absence of novae but dynamically stable in the presence of novae is
quite large, and may be populated by a non--negligible fraction of the
observed CVs. This is illustrated in Figure~1 (right panel) where we
plot the 
critical lower limit on the WD mass for dynamically stable mass
transfer, both for 
the case of conservative mass transfer (upper curve) and for mass
transfer with ballistic nova ejections (lower curve). The dashed lines
(from top to bottom) indicate the lower mass limit on C/O WDs, the
upper mass limit on He WDs and the lower mass limit on He WDs as
obtained by standard population synthesis calculations (e.g.\ Willems,
Mundin \& Kolb, this volume).

Any CVs residing in between the two stability curves contribute to a
significant widening of the mass transfer spectra if condition
(1) is met. 
The overall impact of nova outbursts on the transfer rate spectra must
be assessed with population synthesis calculations. 
As an illustraton, Figure~2 shows results from a standard CV model
population, i.e.\ with standard assumptions about the common envelope
evolution and magnetic braking strength, following the method used by
Howell, Rappaport \& Nelson 2001. The solid curves refer to a
population with constant ejecta mass $M_{\rm ej} = 3 \times 10^{-4}
M_\odot$, while the dashed curves represent a population where the
transfer rate was averaged over individual nova outbursts.

\begin{figure}
\plotfiddle{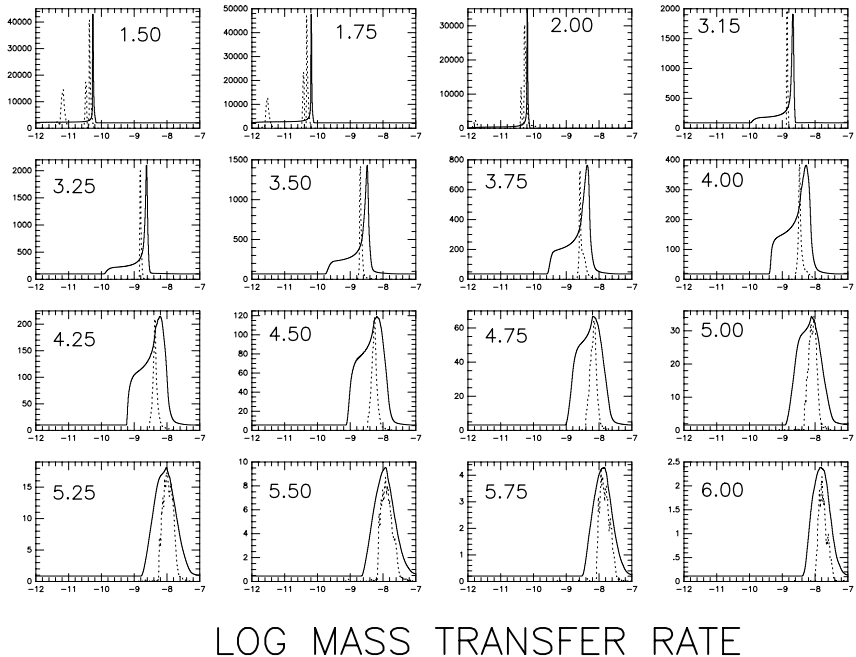}{9cm}{90}{50}{50}{190}{-20}
\caption{Mass transfer spectra at different orbital periods (as
labelled; $P$/hrs). Solid curves: nova ejecta mass $3 \times 10^{-4}
M_\odot$. Dashed: averaged model (no novae).}
\end{figure}

The figure shows that the transfer rate spectra with novae are wider
than in the averaged model for periods in the range $3 \la P/hr \la 5$,
while the differential effect is negligible everywhere else. The mean
value of the 
two different distributions is necessarily the same and set by the AM
loss 
rate, while the width of the nova--affected distributions is
insensitive to the adopted systemic AM loss rate. In other words, the 
effective widening of the spectra works also for a much less efficient
magnetic braking rate.

A higher weighting of brighter (higher $\dot M$) systems would skew
the distributions towards larger values of $\dot M$. In that case the
observed average transfer rate of the weighted distribution
immediately above the gap is larger than the secular mean rate set by
the AM loss rate. This could alleviate the problem that the 
transfer rate at the upper edge of the gap required to 
explain the width of the period gap is smaller than
the critical rate for disc instability, in apparent conflict with the
observed underabundance of dwarf novae in the $3-4$~hr period interval
(e.g.\ Shafter 1992).  

At longer orbital periods the effect of novae is
negligible. Podsiadlowski, Han \& Rappaport (2001) point out that 
the relatively large fraction of CVs with a somewhat evolved
donor star could give rise to a greater width of the mass transfer
spectra above 5~hrs than the ones shown in Figure~2.

\section{CVs on the edge}

The shape of the CV period distribution below 2 hrs continues to
challenge the standard model of CV evolution. The well known
short--period cut--off at $\simeq 78$~min is thought to represent the
period minimum that systems with hydrogen--rich donor stars reach when
their evolution is driven by gravitational radiation (e.g.\ Paczynski
1981, Kolb \& Baraffe 1999, and references therein). The problems
with this interpretation are summarised in Figure~3.
The upper panel shows the lower end of
the CV period distribution ($P<2$~hrs; data from an updated version of
Ritter \& Kolb 1998; see also Downes et al.\ 2001). Even with the most
up--to--date input physics which is so successful in 
describing very low--mass stars and brown dwarfs (e.g.\ Chabrier \&
Baraffe 2000) the calculated value for the minimum period $P_{\rm
min}$ is more than $10$~min smaller than the observed cut--off (middle
panel), 
consistent with earlier findings (e.g.\ Kolb \& Ritter 1992). Perhaps
even more seriously, the very fact that systems
evolve through a state with $\dot P = 0$ implies that there will be an
accumulation of systems at the period minimum (a ``period
spike''). The lower panel of  
Figure~3 shows the discovery probability $\propto 1/\dot P$ for an
ensemble of CVs that form continuously with $0.21 M_\odot$ secondary
mass and $0.6 M_\odot$ WD mass.

\begin{figure}
\plotfiddle{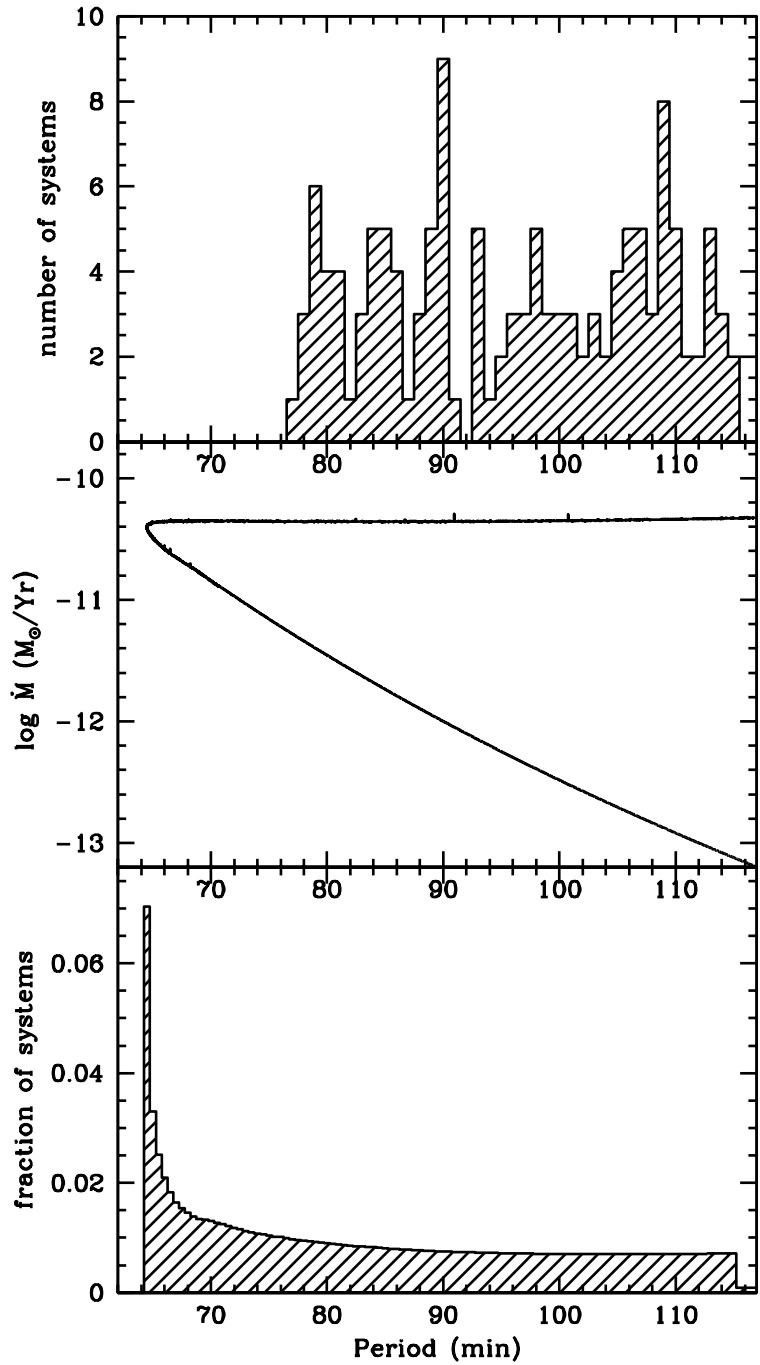}{18cm}{0}{100}{100}{-180}{-170}
\caption{The minimum period problem. Upper panel: distribution of
observed orbital periods below 2 hrs. Middle panel: evolutionary track
of a system where mass transfer is driven by gravitational
radiation. Lower panel: 
corresponding discovery probability.} 
\end{figure}

Various mechanisms have been proposed that would move the calculated
period minimum $P_{\rm min}$ to longer periods. Most assume an
additional form of 
AM losses, e.g.\ magnetic braking. Numerical experiments
show that a rate of roughly three times the rate given by
gravitational radiation (GR) alone is required to raise the calculated
$P_{\rm min}$ to the observed value (Kolb \& Baraffe 1999). The
magnetic braking strength claimed by Andronov et al.\ 2001 is about two
times the GR rate. Barker \& Kolb (2002; see also this volume) show
that consequential AM losses such as accretion disc winds or a
magnetic propeller would raise the period minimum by $\la
5$~min. Increased AM loss is preferred by Patterson (2001), as it
both avoids 
an overpopulation of the ``CV graveyard'' --- systems with a brown
dwarf donor star that have evolved past the period minimum and
continue to transfer mass at a very low rate --- and removes a
claimed systematic discrepancy in the mass--radius relation of 
low--mass single stars and donors in short--period CVs.

Other attempts to reconcile the calculated and observed minimum period
focussed on the Roche model and the approximation of the lobe--filling
component in a semi--deatched binary by a one--dimensional stellar
model. Nelson et al.\ (1985) used corrections for rotational and tidal
deformations of the donor star (Chan \& Chau 1979) and found
a differential increase in $P_{\rm min}$ by 10\%. However, more
recently Kolb \& Baraffe (1999) obtained a much smaller differential
effect ($\simeq 1$ min) using the same prescription but up--to--date
stellar models. 

Using a polytropic equation of state, Rezzolla et al.\ 2001
constructed sequences of self--consistent three--dimensional
hydrostatic models of binaries to investigate the quality of the Roche
approximation 
where the component stars are treated as point masses rather than
extended mass distributions. 
The volume--equivalent radius of the critical lobe, the orbital
AM, and the GR AM loss rate found from the self--consistent models
agree with those from the Roche model to within 1-2\%.  
They also confirmed the validity of the commonly used approximation
for the Roche lobe radius in units of the orbital separation given by
Eggleton (1983). 

As an independent check, Renvoiz\'e et al.\ (2001) calculated
three--dimensional polytropic stars using SPH models with 20,000
particles. Significantly, the volume--equivalent radius of the star 
at the start of mass loss through the $L_1$ point (as depicted in
Figure~4) is {\em larger} than the radius of the same star when it is
non--rotating and in isolation. The radius increase depends on the
polytropic index $n$ and is 6\% for $n=1.5$ (as appropriate for 
fully convective stars near $P_{\rm min}$) and $>10\%$ for
$n=3$, the case shown in the Figure. 

\begin{figure}
\plotfiddle{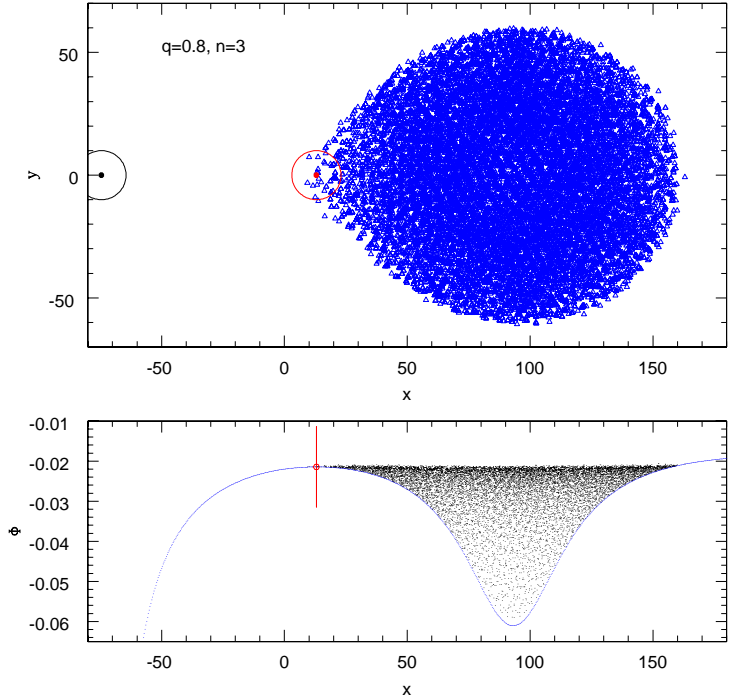}{10cm}{0}{50}{50}{-150}{-70}
\caption{SPH model of the donor star in a semi--detached binary with
mass ratio 0.8. The polytopic index is 3. Upper panel: view of the
orbital plane. The accretor and the saddle point of the effective
potential are encircled. Lower panel: The effective potential along
the line joining the stellar centres.}
\end{figure}

A simple rescaling of the radius of one--dimensional models by a
factor 1.06 increases the calculated minimum period of CVs that are
driven by GR by about 5 mins ($P_{\rm min} \simeq 71$~min if 
$M_1 = 0.6 M_\odot$). This is still well short of the cut--off
period; a radius increase of almost $20\%$ is required to match up
$P_{\rm min}$ with the cut--off (e.g.\ Barker \& Kolb
2002). However, it is not straightforard to include the findings from
the SPH star into one--dimensional stellar models. The expanded star 
is likely to have a lower luminosity and hence a longer thermal
time. This in turn could lead to period bounce at a longer
period. Such a second--order effect cannot be modelled by a simple
rescaling.  

None of the considerations presented so far address the problem of the
missing period spike at $P_{\rm min}$. Barker \& Kolb (2002)
calculated models that effectively ``smear out'' the period spike,
e.g.\ by allowing a mixed population with consequential AM losses of
varying efficiency, or with donor stars that are subject to different
degrees of bloating, e.g.\ due to the effect of magnetic pressure
inside the donor star (as suggested by D'Antona 2000). They find 
that none of the theoretically calculated period distributions fits
the observed one as well as a distribution that is simply flat in $P$.  
It is significant that the missing period spike affects both
magnetic and non-magnetic CVs alike, so it seems unlikely that the
cause should be related to effects in the accretion disc.

The radical hypothesis put forward by King \& Schenker (2002) to
abandon the identification of the cut--off with the minimum period
altogether 
may appear to solve both problems, the mismatch and the missing spike,
in one stroke. The suggestion that the cut--off is due to an age limit
of the population implies that observed CVs close to 80 min orbital
period are still evolving to smaller periods. The age limit hypothesis
appears to reconcile current 
stellar models for short--period CVs that are driven by GR with the
observed period distribution. The same models then tell us that
systems that are currently at 78 min orbital period will continue to
decrease $P$ for another 0.8 Gyr or so, before they reach the minimum
period at about 70 min. This is a measure of how finely tuned the age
limit has to be. More importantly, the mass of {\em all}
hydrogen--rich donor stars in CVs that adhere to the age limit {\em
must} be larger than $\simeq 0.09 M_\odot$. Although there is no CV
known where we have unambiguous proof that the donor mass is smaller
than this limit, there are several excellent candidates where this is 
likely, most notably WZ Sge (e.g.\ Patterson 1998, but see
Steeghs et al.\ 2001). 
The quest to identify brown--dwarf donor CVs should continue; the
first few found would make the age limit hypothesis history.

\section{Conclusions}

This is an exciting era, challenging our understanding of the
evolutionary state of CVs and X--ray binaries. Old paradigms will have
to be re--assessed. The magnetic braking of rapidly rotating 
young, single stars seems to be different from the braking that old,
rapidly rotating CV secondaries experience. The best stellar models to
date continue to ignore the observed period minimum, and despite an
ever increasing number of CVs with determined orbital period they
still refuse to pile up at 80 minutes. Do we really understand
the long--term evolution of CVs? The answer must be a clear
{\em no}.

\bigskip
\acknowledgements{I thank Isabelle Baraffe, John Barker, Steve Howell,
Andy Norton and Saul Rappaport for help in compiling this review.
}

\end{document}